\documentclass[a4paper,fleqn]{cas-sc}

\usepackage[authoryear]{natbib}
\usepackage[version=4]{mhchem}

\usepackage{lineno} 
\usepackage{setspace} 

\def\tsc#1{\csdef{#1}{\textsc{\lowercase{#1}}\xspace}}
\tsc{WGM}
\tsc{QE}
\tsc{EP}
\tsc{PMS}
\tsc{BEC}
\tsc{DE}

\begin{document}
\let\WriteBookmarks\relax
\def\floatpagepagefraction{1}
\def\textpagefraction{.001}


\shorttitle{A Full-Atmosphere Model of Jupiter}

\shortauthors{A Kn\'{i}\v{z}ek et~al.}

\title [mode = title]{A Full-Atmosphere Model of Jupiter}                      
\tnotemark[1]


%
\author[1,2]{Anton\'{i}n Kn\'{i}\v{z}ek}[orcid=0000-0002-0932-0380]



\credit{Conceptualization, Model development, Writing of the paper}

\affiliation[1]{organization={J. Heyrovsk\'{y} Institute of Physical Chemistry, Czech Academy of Sciences},
    addressline={Dolej\v{s}kova 2155/3}, 
    city={Prague},
    postcode={CZ18223}, 
    country={Czechia}}

\affiliation[2]{organization={Atmospheric, Oceanic and Planetary Physics, University of Oxford}, 
addressline={Parks Road}, 
city={Oxford}, 
postcode={OX1 3PU}, 
country = {United Kingdom}}

\author[3]{Paul B. Rimmer}[orcid=0000-0002-7180-081X]
\cormark[1]
\ead{pbr27@cam.ac.uk}
\credit{Conceptualization, Supervision}

\affiliation[3]{organization={Cavendish Laboratory, University of Cambridge},
    addressline={JJ Thomson Ave}, 
    city={Cambridge},
    postcode={CB3 0HE},
    country={United Kingdom}}

\author[1]{Martin Ferus}[orcid=0000-0003-4008-2920]
\credit{Funding}

\cortext[cor1]{Corresponding author}

\begin{abstract}
This paper presents a combined 1D photochemical-thermochemical kinetics model of Jupiter's deeper atmosphere, troposphere and stratosphere.
The model covers atmospheric pressure range from $1.1 \times 10^{3}$~bar to $7.4 \times 10^{-11}$~bar and is the first model that incorporates sulfur chemistry when spanning an  atmospheric region of this extent.
This model incorporates a new version of the STAND reaction network with updated \ce{NH4SH} chemistry, and updated Antoine equation parameters for \ce{NH4SH} and \ce{H2S}.
Validation against current models of Jupiter's atmosphere as well as recent observational data shows that our model successfully describes Jupiter's main observed chemical features.
Since one of the focuses of the model is the chemistry on nitrogen, it correctly predicts the formation of a mixed \ce{NH3}-\ce{NH4SH} cloud layer between 0.1 and 1 bar.
It also describes the chemistry of \ce{HCN} throughout the atmosphere and discovers a region in the stratosphere between $1 \times 10^{-6}$ and $6.76 \times 10^{-8}$~bar, where \ce{HCN} forms through radical chemistry with maximum mixing ratio 33~ppb at $2.94 \times 10^{-7}$~bar -- a prediction testable by observations.
At the same time, our model predicts a quenched \ce{N2} mixing ratio 490~ppm up to 10$^{-6}$~bar.
The model therefore successfully bridges the gap between existing models of separate regions of Jupiter's atmosphere and makes new testable predictions of several chemical species.
\end{abstract}




\maketitle


\section{Introduction} \label{sec:intro}
Jupiter is the largest planet of our Solar system, a failed star, and as such has always attracted considerable attention of the scientific community.
Its atmosphere is mainly composed of hydrogen and helium with no clear lower boundary~\citep{Guillot1999}.
In addition to the two main gases, the atmosphere also contains water, methane, ammonia, hydrogen sulfide and nitrogen.

Throughout the history of studying Jupiter, observations of its atmosphere and modelling were used together to provide a picture of the gas giant in an ever-increasing detail.
In recent years, the importance of understanding Jupiter's chemistry has grown, because the planet represents a class of exoplanets known in extrasolar systems~\citep{Gratton2023}.
Naturally, the most efficient way of collecting planetary data is sending probes to observe the planet from close-up.
The first of these for Jupiter was the Voyager 1 spacecraft, which made a flyby in 1979 and measured the temperature profile, magnetic field and observed its moons.
Recent reanalysis by~\citet{Gupta2022} shows that thermal data from this probe are still valid today.

Another important and successful probe was the Galileo probe, launched in 1989, which was dedicated to analysis of the Jupiter's atmosphere and moons.
Models preceding its launch include the work of~\citet{Prinn1977}, who propose that the presence of CO in what they call the 'visible' atmosphere is due to rapid mixing from deeper atmospheric layers with temperature reaching 1100 K.
Furthermore, ~\citet{Prinn1981}, who proposed nitrogen and hydrogen reactions in Jupiter's atmosphere as an observable prediction for the probe.
That theoretical paper studied reaction kinetics and aerosol catalysis to suggest that nitrogen should be stable in the deep atmosphere of Jupiter and should manifest in the 'upper visible regions' in abundances 0.6 - 3 ppmv without catalysis or 10 ppmv with catalysis considered.
The probe later provided a range of new information on the giant planet, including the atmospheric pressure and temperature (\textit{pT}) profile~\citep{Seiff1998} and built, alongside Earth-based observations, the foundation for many studies dealing with Jupiter's chemistry.
These latter studies included photochemistry and carbon chemistry~\citet{Moses2005}, Jupiter's stratospheric chemistry with focus on carbon species~\citep{Moses2003}, \ce{PH3} and \ce{NH3} photochemistry~\citep{Visscher2009}, or the chemistry of \ce{HCN}~\citep{Moses2010}.
Hydrogen cyanide itself has received significant attention, e.g. in~\citet{Tokunaga1981, Lellouch2006}, and lately by~\citet{Cavalie2023} who observed \ce{HCN} and \ce{CO} with the ALMA telescope and retrieved spatial distribution of those species on Jupiter.

One significant event in the history of Jupiter was the impact of comet Shoemaker-Levy 9 in 1994, which marked the first observation of an extraterrestrial collision in the Solar system~\citep{Crawford1994,Zahnle1994}.
The impact altered the chemical composition of the surrounding atmosphere for years to come, manifested e.g. through new sulfur chemistry~\citep{Moses1995}, and altered \ce{H2O} and \ce{CO2}~\citep{Bezard2002}, or CO, \ce{CS2} and COS~\citep{Lellouch1995,Lellouch1997} mixing ratios.

The most recent wave of information on Jupiter's composition and chemistry was provided by the Juno spacecraft.
For example, observations by~\citet{Bolton2017, Li2017} show variable ammonia levels across the planet.
The formation of mushballs and subsequent rainout have been proposed as a possible explanation~\citep{Guillot2020a, Guillot2020b, Grassi2020, Zhang2020, Ingersoll2017, Stevenson2020,Brown2018}.
Ammonia itself is an important precursor of \ce{NH4SH}, which is expected to form on Jupiter and likely contributes to its orange colour~\citep{Loeffler2018}.
Its presence was also suggested in the Giant Red Spot by~\citet{Bjoraker2018}.
Juno has also significantly improved knowledge of Jupiter's gravity, leading to the idea that Jupiter may have a diluted central concentration of heavy elements.
The mission has also revealed wind depth in the odd zonal harmonics of the gravity field and details about the magnetic field~\citep{Stevenson2020}.

Combination of Juno, Galileo, Voyager and ground-based observational data provides constraints on the structure of the zonal jets~\citep{Kaspi2020}, water abundance~\citep{Li2020}, carbon species abundance~\citep{Sinclair2017, Sinclair2018}, depletion of alkali metals~\citep{Bhattacharya2023}, or the abundance of oxygen~\citep{Cavalie2023b}.
Such information can be used for example to constrain the formation location of Jupiter, which according to~\citet{Oberg2019} formed beyond the \ce{N2} snowline and then migrated towards the Sun.

Still more is expected from the upcoming JUICE mission in the 2030s~\citep{Fletcher2023}.
The main topics of this mission will include acquiring a global perspective on Jupiter's atmospheric science, investigation of phenomena on timescale from minutes to months, and exploration of the atmosphere at high vertical resolution.
The mission will also help answer how the atmosphere is connected to the deep circulation and composition, which is currently an aspect of the Jupiter's atmospheric research that relies heavily on assumptions about the deeper atmosphere.
Yet, it is at the same time understood that for example the distribution of ammonia is tied to deep atmosphere dynamics~\citep{Stevenson2020} or that the distribution of carbon species is heavily dependent on the deep water content~\citep{Visscher2010}.

Most models of the Jupiter's atmosphere focus on specific questions, such as the behaviour of a single species, constraint on a single parameter, or specific region.
Some notable works include~\citet{Ingersoll1969}, who showed that the light and dark bands on Jupiter have different temperatures.
\citet{Fegley1994} then assumed thermochemical equilibrium and a dry adiabatic thermal profile, and constructed a model of Jupiter and Saturn without any probe data.
In the same year,~\citet{Guillot1994} challenged the view that giant planets are adiabatic and fully convective below 4000 K.
The authors later reevaluated their work~\citep{Guillot2006}, and as discussed in~\citet{Freedman2008}, the use of correct pressure-broadened alkali opacities (not known at the time of the original publication) leads to the conclusion that the original fully convective hypothesis still holds, which goes on to show that the use of up-to-date data and models is vital for successful description of planetary atmospheres.
It is nowadays generally recognized that Jupiter's deeper atmosphere is in thermodynamic equilibrium.
Many studies devoted to interior studies were performed, one of the latest being~\citet{Rensen2023}.
There, Juno data were used to model elemental and molecular abundances in the deeper atmosphere.
Other models, such as~\citet{Miguel2022} or~\citet{Howard2023} reveal important information about the distribution of some heavier elements and the structure of the planetary interior.
These papers use thermodynamic equilibrium codes, such as GGchem~\citep{Woitke2018}.
A significant advantage of these codes is the usage of Gibbs free energy minimization to calculate the equilibrium composition based on thermodynamic properties of selected species.
These calculations are fast and efficient and allow for the exploration of a large parameter space quickly.

On the other hand, as one progresses upwards, the atmosphere quickly departs from equilibrium, and kinetics and photochemistry become an important aspect~\citep{Moses2014}.
For example,~\citep{Moses2011} discusses carbon, nitrogen and oxygen chemistry on two hot Jupiters and shows that disequilibrium processes enhance the presence of \ce{CH4} and \ce{NH3} above equilibrium due to quenching.
Both species are then photochemically removed in the upper atmospheric layers.
Such disequilibrium chemistry also enhances the mixing ratios of unsaturated hydrocarbons (mainly \ce{C2H2}), nitriles (HCN in particular), and radicals (OH, \ce{CH3}, \ce{NH2}).
Similarly,~\citet{Hu2021} shows that photodissociation of \ce{NH3} in presence of \ce{CH4} on temperate gas giant planets leads to increased mixing ratio of \ce{HCN}.
At the same time, photodissociation of \ce{CH4} in the presence of \ce{H2O} leads to CO and \ce{CO2} and suppresses the synthesis of \ce{HCN}.
They also show that \ce{NH3} is photochemically removed from Jupiter's atmosphere and once turned to \ce{N2H4}, the vast majority condenses out.
The effect of vertical winds and eddy diffusion on the distribution of ammonia is explored in another disequilibrium model of Jupiter's deeper atmosphere and stratosphere that covers the same atmospheric extent as our model \citep{Tsai2021}.

What all these studies show is that disequilibrium processes are very important in atmospheres of giant planets.
Moreover, lack of direct observations from the deeper atmosphere further cements the need for correct understanding of both thermochemistry and chemical kinetics in planetary atmospheres as deeper abundances of species are often constrained by observations in the upper atmospheric layers and subsequent modelling~\citep{Visscher2010, Wang2015, Wang2016}.

In this paper, we employ a 1D photochemical-thermochemical kinetics model ARGO to describe the chemistry in equilibrium and the disequilibrium processes in a combined code (Section \ref{sec:chemical_model}).
The model constructed in this paper is intended to bridge the existing gap between the deeper and upper regions of Jupiter's atmosphere and provide a tool for the prediction of observable features in Jupiter's atmosphere, many of which depend on deeper atmospheric properties.
The implemented gas phase chemistry is described in Section~\ref{sec:gas_phas_chemistry}, the ammonia and ammonium hydrosulfide are shown in Section~\ref{sec:NH3_NH4SH}, the deep tropospheric chemistry is discussed in Section~\ref{sec:deep troposphere}, and the chemistry of HCN is described in Section~\ref{sec:HCN}.

\section{Chemical model} \label{sec:chemical_model}
The basis of the code ARGO used in this paper was extensively described in~\citet{Rimmer2016}.
In brief, it is a 1D photochemical-thermochemical code for atmospheric chemistry modelling.
The code combines photochemistry, kinetics, diffusion, condensation, evaporation and outgassing. The thermochemical portion of the code includes reverse reaction rate coefficients following a thermodynamic Gibbs free energy treatment.
The code was described in detail in \citep{Rimmer2016}.
Necessary inputs for this code include:
\begin{itemize}
    \item combined pressure (\textit{p}, bar), temperature (\textit{T}, K), and eddy diffusion coefficient ($K_{\mathrm{zz}}$, cm$^{2}$\,s$^{-1}$) profile
    \item initial atmospheric composition at one of the boundaries, here at the bottom of the atmosphere
    \item actinic flux at the top of the atmosphere
    \item chemical network
\end{itemize}
At each atmospheric level, the code solves a set of coupled 1D continuity equations which describe the time-dependent atmospheric composition at that level:
\begin{equation}
    \frac{\delta n_{i}}{\delta t} = P_{i} - L_{i} -\frac{\delta\Phi_{i}}{\delta {z}}
\end{equation}
where $n_{i}$ [cm$^{-1}$] are number densities of chemical species in the atmosphere, $t$ is time, $P_{i}$ [cm$^{-3}$\,s$^{-1}$] are the production rates and $L_{i}$ [cm$^{-3}$\,s$^{-1}$] the destruction rates, $z$ is the altitude and $\Phi$ is the flux of species $i$.
This flux includes both eddy ($K$ [cm$^{-2}$\,s$^{-1}$]) and molecular ($D$ [cm$^{-2}$\,s$^{-1}$]) diffusion. 
Implementation of eddy diffusion and molecular diffusion is described in \citep{Rimmer2016} in detail.

\subsection{Pressure-temperature profile} \label{subsec: pT_profile}
In this paper, we use a combined (\textit{p,T}) profile, where the upper part from 6.7 bar (altitude -63.7 km) to the top is taken from~\citet{Moses2005}.
That profile reasonably agrees with a profile published by~\citet{Seiff1998}, who presents data obtained by the Galileo probe.
The lower part of the (\textit{p,T}) profile for altitudes below -63.7 km was calculated according to~\citet{Visscher2010}, whose approach stems from~\citet{Fegley1985}.
The bottom of the profile temperature was chosen to be 1400 K, because below that the code has convergence issues and the atmosphere follows equilibrium.
First, the dry adiabatic lapse rate was used to calculate the altitude grid for the profile:
\begin{equation}
    \frac{\delta T}{\delta z} = \frac{Mg}{C_{\mathrm{p}}}
\end{equation}
where $T$ is the temperature, $z$ is the altitude, $g$ = 24.79~m\,s$^{-2}$ is the surface gravity on Jupiter, $C_{\mathrm{p}}$ is the specific heat capacity of the atmosphere and $M = 2.33$~g\,mol$^{-1}$ is the average molar mass of the atmosphere.
For the calculation, $g, M$ and $C_{\mathrm{p}}$ were all considered as independent on the altitude and composition of the atmosphere.
The specific heat capacity was estimated as:
\begin{equation}
    C_{\mathrm{p}} = c_{\mathrm{p}, \mathrm{H}_{\mathrm{2}}}x_{\mathrm{H}_{\mathrm{2}}} + c_{\mathrm{p}, \mathrm{He}}x_{\mathrm{He}}
\end{equation}
where $c_{\mathrm{p, H_{2}}} = 28.836$~J\,K$^{-1}$\,mol$^{-1}$ and $c_{\mathrm{p, He}} = 20.786$~J\,K$^{-1}$\,mol$^{-1}$ are the specific heat capacities of \ce{H2} and \ce{He}, respectively, and $x_{\mathrm{H_{2}}} = 0.86219$ and $x_{\mathrm{He}} = 0.136$ are the mixing ratios of \ce{H2} and \ce{He} taken from~\citet{Moses2005} at 6.78~bar.
The \ce{H2} and \ce{He} are the dominant species in the atmosphere and the calculated specific heat capacity should be reasonably accurate.
Then, the pressure profile was calculated according to~\citet{Seiff1998} using:
\begin{equation}
    log_{10}\left(\frac{p_{n+1}}{p_n}\right) = \frac{C_\mathrm{p}}{R}log_{10}\left(\frac{T_{n+1}}{T_{n}}\right)
\end{equation}
where $p$ stands for pressure and $R = 8.314$~J\,K$^{-1}$\,mol$^{-1}$ is the universal gas constant.
The profile has a vertical resolution 1 km below the Galileo data and 637 levels in total.

\begin{figure}[ht!]
    \centering
    \includegraphics{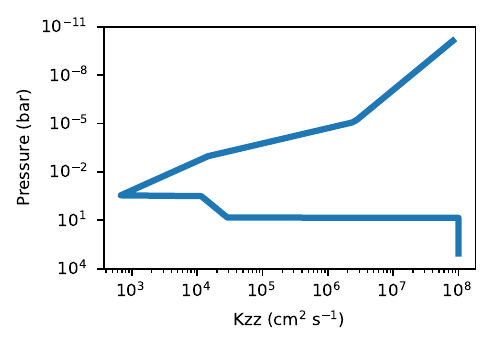}
    \caption{Eddy diffusion coefficient adopted in our work, units cm$^2$ s$^{-1}$.}
    \label{fig:Kzz}
\end{figure}

The eddy diffusion coefficient, $K_{\mathrm{zz}}$ for the stratosphere was also taken from~\citet{Moses2005}.
For the deeper atmosphere, we use the value of~\citet{Visscher2010}, who estimate a value $1\times10^8$~cm$^2$\,s$^{-1}$ which is altitude-independent.
They also test the sensitivity of their model to changes in the value of $K_{\mathrm{zz}}$ and even though differences arise, the original value remains a valid approximation.
The adopted Kzz profile is shown in Figure \ref{fig:Kzz}.

\begin{figure}[ht!]
    \centering
    \includegraphics{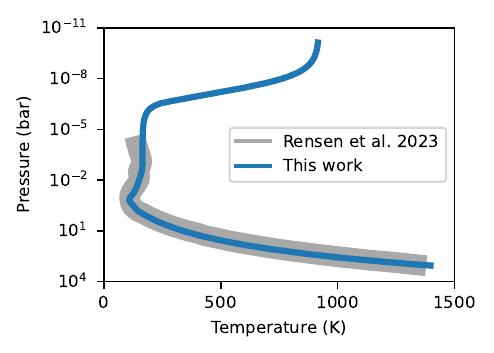}
    \caption{Comparison of (\textit{p,T}) profiles used in this work and in~\citet{Rensen2023}. Note that our profile above 6.7 bar was adopted from \citet{Moses2005}.}
    \label{fig:pT_profile_comparison}
\end{figure}

Our (\textit{p,T}) profile was also compared to the profile used recently by~\citet{Rensen2023}.
In that paper, the authors an adiabat which extends from the Galileo data to the deep interior.
The profile was calculated by~\citet{Miguel2022} using the CEPAM code~\citep{Guillot1995}.
Figure \ref{fig:pT_profile_comparison} shows that both approaches yield comparable results.

\subsection{Initial conditions} \label{subsec: init_cond}
The initial conditions to this model are supplied as elemental abundances at the bottom of the (\textit{p,T}) profile.
The elemental abundances of H, He, N, C, O and S were taken from \citet{Rensen2023}, who uses the abundances of \ce{CH4} and \ce{H2S} from \citet{Wong2004} and \ce{H2O} and \ce{NH3} from~\citet{Li2020}.
Their model uses the lower boundary $\sim$1000 K, similar to our model, and so this usage is justified.
Ar abundance was taken from~\citet{Visscher2010}, because~\citet{Rensen2023} did not include Ar in the model.
The abundances were converted to mixing ratios and scaled to 1.
We also note that~\citet{Rensen2023} take values from various sources (including molecular abundances) and recalculate them to elemental abundances, but do not list errors, and the original information is lost during their conversion.
We still choose to use their values because they are to most up-do-date values.
Therefore, to test the sensitivity of the calculation, an artificial 5\% margin on each elemental abundance was introduced instead.
Since ARGO uses mixing ratios as initial conditions, the 5\% margin was applied to the mixing ratios directly, i.e., $n_{i,max} = n_i*1.05$, etc., where $n_i$ is the mixing ratio of species \textit{i} and $n_{i, max}$ is the mixing ratio of that species + 5\% margin.
This does not correspond to 5\% change in metallicity.
The elemental ratios are shown in Table \ref{Table_init_cond}.
The table also shows the values in log units vs H as is customary in astronomy (for definition see~\citet{Asplund2009}) and enrichment vs solar calculated from our values vs solar elemental abundances from~\citet{Asplund2009}.
The calculated values compare Jupiter's deep atmosphere vs solar photosphere and therefore differ from what is commonly observed in the literature (e.g.~\citet{Oberg2019}).

\begin{table}[width=.9\linewidth,cols=4,pos=h]
    \caption{The table shows elemental abundances (mole fractions) at the bottom of the atmosphere used in our model. These values are also shown in log units vs H (as customary in astronomy). Enrichment vs solar abundances from~\citet{Asplund2009} was also calculated for these values.} \label{Table_init_cond}
    \begin{tabular*}{\tblwidth}{@{} LLLL@{} }
        \toprule
        Element & Mole fraction & Elem. abund. & Enrichment \\
        \midrule
        \ce{H} &  $9.0726\times10^{-1}$ &  12.00 &  1.00 \\
        \ce{He} & $8.6642\times10^{-2}$ &  10.98 &  1.12 \\
        \ce{C} & $2.2080\times10^{-3}$ &  9.37 &  9.04 \\
        \ce{N} & $4.6813\times10^{-4}$ &  8.71 &  7.63 \\
        \ce{O} & $3.3176\times10^{-3}$ &  9.56 &  7.46 \\
        \ce{S} & $9.0947\times10^{-5}$ &  8.00 &  7.60 \\
        \ce{Ar} & $1.6512\times10^{-5}$ &  7.26 &  7.24 \\
        \bottomrule
    \end{tabular*}
\end{table}

We have also compared our code to FastChem~\citep{FastChem} and GGchem~\citep{Woitke2018} and found agreement in atmospheric composition at the bottom of our model.
The only notable difference is a different abundance (<10\% difference) of the two main nitrogen-bearing species -- \ce{N2} and \ce{NH3}, which is likely caused by the fact that not all reactions involving those species are reversed in ARGO.
Specifically, ARGO does not at all include reverse three-body reactions that involve three trace species, because the probability of such reactions happening is very low.
An example of such reaction would be for example
\begin{align}
    &\ce{NH + H -> N + H + H} \\
    &\ce{NH + NH -> N2 + H + H}
    \label{eq:non-reversed_example}
\end{align}

While the equilibrium codes use the law of mass action, element conservation equations, and Gibbs free energy minimization to calculate chemical equilibrium composition, ARGO uses kinetics with reaction rates extrapolated to these high temperatures.
Importantly, ARGO, too, reaches chemical equilibrium at this atmospheric level, as observed from the fact that all abundances at the bottom of the model are almost uniformly vertically mixed. This is not a grand coincidence! The forward and reverse rates are balanced by equilibrium constants, which in turn are set by the Gibbs free energy of the reaction. It can be shown that this approach will reproduce equilibrium precisely, if units are properly accounted \citep{Visscher2011}.
At the same time, not all species present in ARGO are included in FastChem and GGchem, which also influences the result of the calculation.
Overall, the differences are not major and do not influence the upper atmospheric results significantly.

\subsection{Actinic flux} \label{subsec: actinic_flux}
The actinic flux at the top of the atmosphere (the total amount of radiation at the top of the atmosphere in units cm$^{-2}$\,s$^{-1}$\,\AA$^{-1}$) was taken from~\citet{Rimmer2016}, who take their profile from~\citet{Huebner1979},~\citet{Huebner1992}, and~\citet{Huebner2015}.
That flux, original for Earth, was scaled to the mean orbital distance of Jupiter ($5.2038$~AU semi-major axis) with the inverse square law.

\subsection{Chemical network} \label{subsec: chem_network}
The chemical network used in this paper is an update to the previously published chemical network from~\citet{Rimmer2021}.
Two existing reaction rates were updated in the network according to~\citet{Hu2021}.
The new rates are shown in Table \ref{tab_new_rates}.

\begin{table}[width=.9\linewidth,cols=2,pos=h]
    \caption{Updated reaction rates with respect to~\citet{Rimmer2021}.} \label{tab_new_rates}
    \begin{tabular*}{\tblwidth}{@{} LL@{} }
        \toprule
        Reaction & Rate (cm$^3$\,s$^{-1}$) \\
        \midrule
        \ce{NH + NH2 -> N2H2 + H} & $1.5 \times 10^{-10} \left(\frac{T}{298}\right)^{-0.27} e^{\frac{38.5}{T}}$ \\
        \ce{NH2 + CH4 -> CH3 + NH3} & $5.75 \times 10^{-11} e^{-\frac{6952}{T}}$ \\
        \bottomrule
    \end{tabular*}
\end{table}

\subsubsection{\ce{NH4SH} chemistry} \label{subsubsec: nh4sh_chem}
Additionally, \ce{NH4SH} formation and destruction were added:
\begin{equation}
    \ce{NH3 + H2S <=>[\text{k$_f$}][\text{k$_r$}] NH4SH}
    \label{eq_NH4SH}
\end{equation}
\ce{NH4SH} was first added to the code as a species.
NASA polynomial coefficients as described in~\citet{Burcat2005} are used for calculating various thermochemical values during the computation.
Examples of these are values is the Gibbs free energy (as $\frac{G_T^{\circ}}{RT}$):
\begin{equation}
\begin{split}
    \frac{G_T^{\circ}}{RT} = & a_1(1-\ln T) - \frac{a_2 T}{2} - \frac{a_3 T^2}{6} \\
    & - \frac{a_4 T^3}{12} - \frac{a_5 T^4}{20} + \frac{a_6}{T} - a_7
\end{split}
\end{equation}
and the reaction equilibrium constant in terms of concentrations, $K_c$:
\begin{equation}
\begin{split}
    K_c = & (RT)^{-\Delta\nu} \exp \bigl(\Delta a_1(1-\ln T) + \frac{\Delta a_2 T}{2} + \frac{\Delta a_3 T^2}{6} \\
    & + \frac{\Delta a_4 T^3}{12} + \frac{\Delta a_5 T^4}{20} + \frac{\Delta a_6}{T} + a_7\bigr)
\end{split}
\end{equation}
where $a_1$ -- $a_7$ are the NASA 7 term polynomial coefficients, $R$ [J mol$^{-1}$ K$^{-1}$] is the universal gas constant, $T$ [K] is temperature, and $\Delta\nu$ is the change in the mole number $\Delta\nu = \sum \nu_j$.
The changes in coefficients are $\Delta a_{ij} = \sum \nu_j a_{ij}$ with the stoichiometric coefficients $\nu_j$ being positive for products and negative for reactants.
In the code, as well as in the database, there exist two sets of polynomial coefficients, one for temperatures up to 1000~K and the other for 1000 -- 6000~K.
There exists a 15\textsuperscript{th} coefficient in the Burcat database, which we do not use.
In our case, we use an expression for the true equilibrium constant, which we define as 
\begin{equation}
\begin{split}
    K = \exp & \left(\Delta a_1(1-\ln T) + \frac{\Delta a_2 T}{2} + \frac{\Delta a_3 T^2}{6} \right.\\
    &\left. + \frac{\Delta a_4 T^3}{12} + \frac{\Delta a_5 T^4}{20} + \frac{\Delta a_6}{T} + a_7 \right)
    \label{Burcat_K_eq}
\end{split}
\end{equation}

The 7 term polynomial coefficients for \ce{NH4SH} that are used in this code are not available.
Instead, we calculate them using the available coefficients for \ce{NH3} and \ce{H2S} and the expression for the equilibrium constant from~\citet{Lewis1969}:
\begin{equation}
    \ln K = 14.82 - \frac{4705}{T}
    \label{Lewis_eq}
\end{equation}
The constant in~\citet{Lewis1969} is valid in the temperature range 180--300~K~\citep{Carlson1987}, and we extrapolate to the whole temperature range.
Also, this expression is derived for the reaction
\begin{equation}
    \ce{NH4SH(s) <--> NH3(g) + H2S(g)}
\end{equation}
In this reaction, \ce{NH4SH} forms in the solid phase.
Our approach adopts that but the code also includes condensation and evaporation (described below) which quickly establish equilibrium between the condensed and evaporated \ce{NH4SH}.
Using the expressions for equilibrium constants in Eqs. \ref{Burcat_K_eq} and \ref{Lewis_eq} above and the 7 term polynomial coefficients for \ce{NH3} and \ce{H2S}, the polynomial coefficients for \ce{NH4SH} were fit.
The final two sets of the 7 term polynomial coefficients are shown in Table \ref{Table_NH4SH_coeffs}.

\begin{table}[width=.9\linewidth,cols=3,pos=h]
    \caption{Calculated NASA 7 term polynomial coefficients for \ce{NH4SH}.}
    \label{Table_NH4SH_coeffs}
    \begin{tabular*}{\tblwidth}{@{} LLL@{} }
        \toprule
        Coefficient & 100--1000 K & 1000--6500 K \\
        \midrule
        $a_1$ & $8.5810$ & $5.0745$ \\
        $a_2$ & $-7.5669\times10^{-3}$ & $9.7451\times10^{-3}$ \\
        $a_3$ & $2.9355\times10^{-5}$ & $-3.2313\times10^{-6}$ \\
        $a_4$ & $-2.7349\times10^{-8}$ & $4.9816\times10^{-10}$ \\
        $a_5$ & $9.0373\times10^{-12}$ & $-2.8795\times10^{-14}$ \\
        $a_6$ & $-1.5095\times10^{4}$ & $-1.4531\times10^{4}$ \\
        $a_7$ & $-1.4632\times10^{1}$ & $1.5549$ \\
        \bottomrule
    \end{tabular*}
\end{table}

The rate constant $k_f$ for the reaction \ref{eq_NH4SH} was derived from the kinetic theory of gases.
From~\citet{Draine2011}, for a reaction
\begin{equation}
    \ce{A + B -> products}
\end{equation}
the reaction rate per unit volume is
\begin{equation}
    k = n_A n_B \langle \sigma v \rangle _{AB}
\end{equation}
where $\langle \sigma v \rangle_{AB}$ is the two-body collisional rate coefficient, and $n_A$ and $n_B$ are number densities of reactants A and B, respectively, in units [cm$^{-3}$].
Then from the kinetic theory of gases, we take
\begin{equation}
    E_r = \frac{3}{2}k_{\mathrm{B}} T
    \label{eq:energy_gas}
\end{equation}
where the energy of an ideal gas is proportional to the temperature of the gas and $k_{\mathrm{B}}$ is the Boltzmann constant.
From~\citet{ChemLibre}, the collisional rate constant can be expressed as 
\begin{equation}
    \langle \sigma v \rangle_{AB} = \langle v_{\mathrm{r}} \rangle \sigma_{AB} e^{-\frac{E_r}{k_{\mathrm{B}} T}}
\end{equation}
where $v_{\mathrm{r}} = \vec{v_{\mathrm{a}}} - \vec{v_{\mathrm{b}}}$ is the relative velocity of the colliding particles, $\sigma_{AB}$ is the averaged sum of hard-sphere collision cross-sections of particles A and B in units [m$^{2}$], $E_r$ is the relative kinetic energy.
Since we consider gas with zero internal energy, the total energy can be expressed as the kinetic energy
\begin{equation}
    E_r = \frac{1}{2}\mu v_{\mathrm{r}}^2
    \label{eq:E_r}
\end{equation}
where $\mu$ is the reduced mass of the reacting particles.
By comparing Eq. \ref{eq:energy_gas} and Eq. \ref{eq:E_r}, we obtain
\begin{equation}
    v_{\mathrm{r}} = \sqrt{\frac{3k_{\mathrm{B}} T}{\mu}}
\end{equation}
and finally
\begin{equation}
    \langle \sigma v \rangle_{AB} = \sqrt{\frac{3k_{\mathrm{B}} T}{\mu}} \sigma_{AB} e^{-\frac{3}{2}}
    \label{rate_kin_eq}
\end{equation}
The molar masses of \ce{NH3} and \ce{H2S}, $M_{\mathrm{\ce{NH_3}}}$ and $M_{\mathrm{\ce{H_2S}}}$ are 17.031~g\,mol$^{-1}$ and 34.08~g\, mol$^{-1}$, respectively.
Hence,
\begin{equation}
    \mu = \frac{M_{\mathrm{H_2S}} M_{\mathrm{NH_3}}}{M_{\mathrm{H_2S}} + M_{\mathrm{NH_3}}} \frac{1}{N_{\mathrm{A}}} = 1.886 \times 10^{-26} \mathrm{kg}
\end{equation}
where $N_{\mathrm{A}}$ is the Avogadro constant.

The kinetic diameters of \ce{NH3} and \ce{H2S} are 260~pm and 360~pm, respectively~\citep{Matteuci2006,Breck1973}, so $\sigma_{AB}$ is then calculated as
\begin{equation}
    \sigma_{AB} = \pi \left(\frac{d_A d_B}{2}\right)^2 = 3.019 \times 10^{-15} \mathrm{cm}^{2}
\end{equation}

A new reaction type was introduced in the code.
The reaction rate is calculated from
\begin{equation}
    k = k_f n_A n_B
\end{equation}
where the reaction rate constant, $k_f$, is calculated as:
\begin{equation}
    k_{f} = c_1 T^{c_2} N_{\mathrm{HTOT}}
\end{equation}
where by comparison with Eq. \ref{rate_kin_eq}
\begin{equation}
    c_1 = \sqrt{\frac{3k_{\mathrm{B}}}{\mu}} \sigma_{AB} e^{-\frac{3}{2}} = 3.16 \times 10^{-12} \mathrm{cm}^3 \mathrm{s}^{-1}
\end{equation}
\begin{equation}
    c_2 = \frac{1}{2}
\end{equation}
and $N_{\mathrm{HTOT}}$ is the number density of the atmosphere which accounts for the fact that the reaction in fact needs a third body to carry away excess energy.

The reverse reaction rate constant is calculated in a similar manner as described in~\citet{Rimmer2016} and~\citet{Visscher2011}.
In short, the expression for the rate constant is
\begin{equation}
    k_r = k_f e^{\frac{-\Delta G_{\mathrm{r}}}{RT}} \left(RT\right)^{\Delta v}
\end{equation}
where $\Delta G_{\mathrm{r}}$ is the Gibbs free energy of the reverse reaction and $\Delta v = \sum_j v_j$, where \textit{j} are the stoichiometric coefficients of the reaction, positive for products and negative for the reactants.

\subsection{Condensation and evaporation} \label{subsec: condens}
The present code also contains a subroutine for condensation and evaporation of various species, which, again was extensively described in~\citet{Rimmer2016, Rimmer2021}.
Evaporation is treated as the exact reverse process of condensation in the code, so all the following discussion pertaining to condensation applies to evaporation as well.
For the purpose of this study, the condensation and evaporation of \ce{H2S} and \ce{NH4SH} were added.
They are based on the Antoine equation and Table \ref{tab_condensation} shows the equations used for \ce{H2S} and \ce{NH4SH} condensation taken from~\citet{Stull1947}.

\begin{table*}[width=.9\textwidth,cols=3,pos=h]
    \caption{Antoine equations used for the condensation of \ce{H2S} and \ce{NH4SH}}.
    \label{tab_condensation}
    \begin{tabular*}{\tblwidth}{@{} LLL@{} }
        \toprule
        Species & Antoine equation & Temperature range (K) \\
        \midrule
        \ce{NH4SH} &
        $\log_{\mathrm{10}}P = 6.09165 - \left(\frac{1598.4}{-43.81+T}\right)$
        & 222.1\footnote{Extrapolated to 110~K in the code.} - 306.4 \\
        \ce{H2S} & 
        $\log_{\mathrm{10}}P = 4.4368 - \left(\frac{8.2944}{-2.5412+T}\right)$
        & 138.8 - 212.8 \\
        \bottomrule
    \end{tabular*}
\end{table*}

As noted in the Table \ref{tab_condensation}, the Antoine equation parameters for \ce{NH4SH} available in the literature are valid between 222.1~K and 306.4~K.
We therefore extrapolate equation to 110 K in order to allow condensation to take place in the Jupiter's atmosphere to the lowest used temperatures.

It is worth mentioning that~\citet{Young2019} use a slightly different approach to the chemistry of \ce{NH4SH}.
The authors, too, start from the mentioned~\citet{Lewis1969} work.
In their global circulation model, they solve for \ce{NH4SH} at each level separately.
They first calculate potential equilibrium between \ce{NH4SH} (s), \ce{NH3} and \ce{H2S}.
Then, potential (would-be) partial pressures ($pp_x$) of \ce{NH3} and \ce{H2S} are calculated as if all \ce{NH4SH} was converted back to vapour.
Next, the code compares these potential partial pressures to the current state of the model and check for
\begin{equation}
    \ln(pp_{\ce{NH3}}*pp_{\ce{H2S}}) > \ln K
\end{equation}
where $K$ is from~\citet{Lewis1969}.
If this condition is satisfied, the code calculates $pp_{\ce{NH4SH}}$ which forms from \ce{NH3} and \ce{H2S} in excess to their saturated vapour pressures.
If the product of vapour pressures of \ce{NH3} and \ce{H2S} is lower than what it would be at the saturated vapour pressures, the \ce{NH4SH} mixing ratio is set to 0.
This is a different implementation based on the same data.
Direct comparison of the results is infeasible since we use a 1D thermochemical/photochemical model with focus on the chemistry of nitrogen and~\citet{Young2019} uses a 3D global circulation model with focus on weather.
Similar approach was also employed by~\citet{Hu2021} as well.

\section{Gas Phase Chemistry} \label{sec:gas_phas_chemistry}
\begin{figure*}
    \centering
    \includegraphics[width=\textwidth]{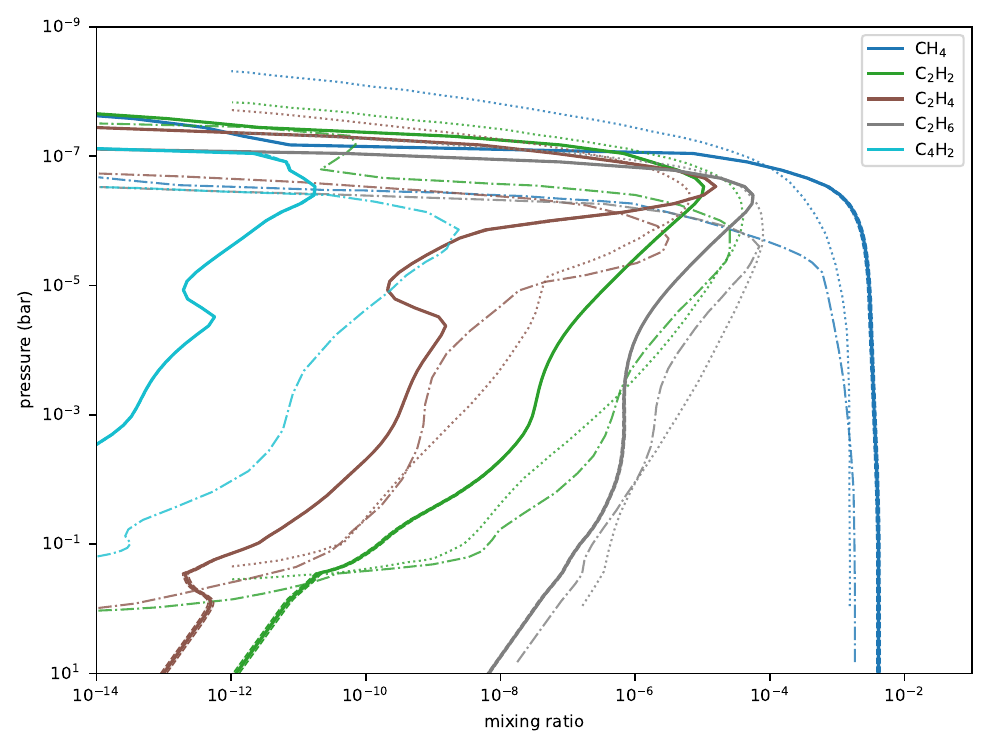}
    \caption{Mixing ratios of \ce{CH4}, \ce{C2H2}, \ce{C2H4}, \ce{C2H6} and \ce{C4H2} as a function of pressure compared to~\citet{Rimmer2016}, Fig. 12, and~\citet{Hu2021}, Fig. 1. Our results are shown in solid line with dashed line margins. The~\citet{Rimmer2016} data are shown in dash-dot line and the~\citet{Hu2021} data are shown in dotted line.}
    \label{fig:J135_as_Rimmer2016_fig_12_vs_Hu2021}
\end{figure*}

After construction, our model was first compared to already existing models to verify its behaviour.
Figure \ref{fig:J135_as_Rimmer2016_fig_12_vs_Hu2021} shows the mixing ratios of \ce{CH4}, \ce{C2H6}, \ce{C2H4}, \ce{C2H2} and \ce{C4H2} (solid line, margins in dashed line), which are compared to the upper atmospheric models of~\citet{Rimmer2016} (dash-dot line) and~\citet{Hu2021} (dotted line).
Our results show 0.4\% \ce{CH4} at 10~bar pressure, which is slightly higher than $\sim$0.19\% of~\citet{Rimmer2016} and~\citet{Hu2021}.
The figure also shows the methane mixing ratio sharply drops off at a higher altitude ($10^{-8}$~bar) as compared to~\citet{Rimmer2016} (where the drop-off is at $\sim 5\times 10^{-7}$~bar), but lower than in~\citet{Hu2021}, where the drop-off is both less sharp and at lower pressures.
Furthermore, mixing ratio of the main observed aliphatic hydrocarbon species (\ce{C2H2, C2H4, C2H6, C4H2}) exhibit similar drop-offs for the model of~\citet{Hu2021}, whereas~\citet{Rimmer2016} predicts both the peak and drop-off at $~10\times$ higher pressure.
Finally,~\citet{Rimmer2016} predict \ce{C4H2} mixing ratios to peak at $10^{-9}$ around the $10^{-6}$ bar level, whereas our updated model predicts the peak at $10^{-11}$ around $10^{-7}$ bar, which is about $100\times$ lower.
\citep{Hu2021} do not include \ce{C4H2} in their model.

\begin{figure*}
    \centering
    \includegraphics[width=\textwidth]{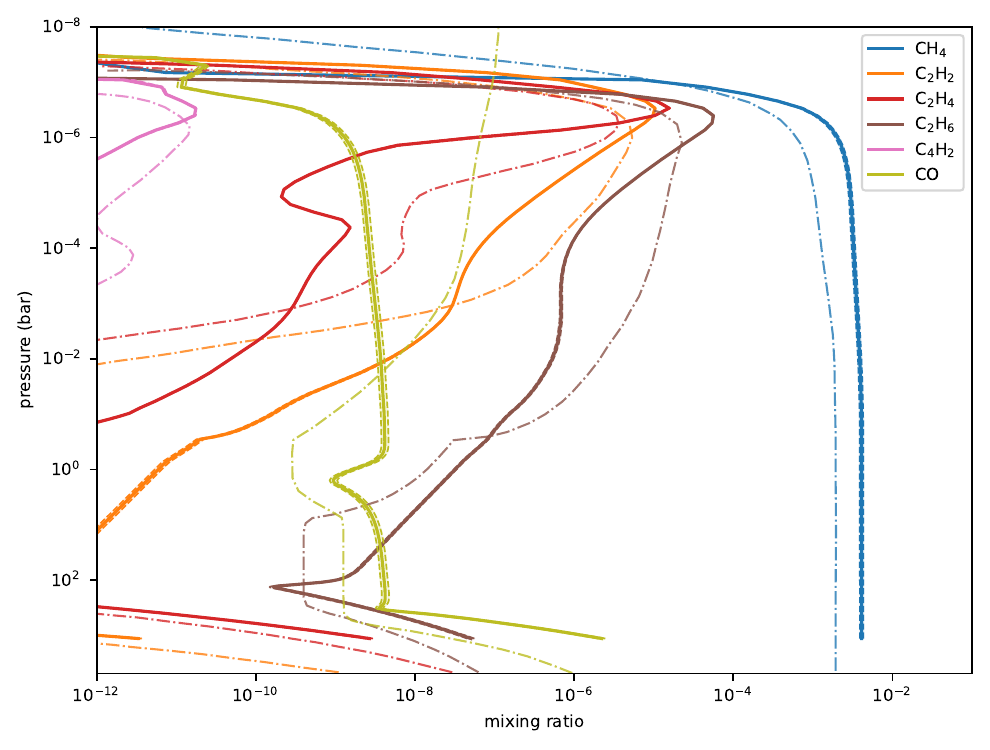}
    \caption{Mixing ratios of \ce{CH4}, \ce{C2H2}, \ce{C2H4}, \ce{C2H6}, \ce{CO} and \ce{C4H2} as a function of pressure compared to~\citet{Tsai2021}, Fig. 15. Our results are shown in solid line with dashed line margins. The~\citet{Tsai2021} data are shown in dash-dot line.}
    \label{fig:J135_as_Tsai2021_fig5}
\end{figure*}

Next, the model was also compared to~\citet{Tsai2021}.
That paper uses the VULCAN code, which falls into the same atmospheric code category ARGO - a disequilibrium code with photochemistry and condensation.
The results are again comparable, as seen from Figure \ref{fig:J135_as_Tsai2021_fig5}.
Especially \ce{CH4} and \ce{C2H6} exhibit similar behaviours.
On the other hand, ARGO produces more \ce{C2H2} and \ce{C2H6} in the deeper atmosphere.
An interesting discrepancy between our models is found the amount of CO, which reasonably agrees below 10$^{-1}$ bar, but above that,~\citet{Tsai2021} predict larger amounts than ARGO.
This is because~\cite{Tsai2021} include a top-of-the atmosphere influx of CO in their model and ARGO does not.
Their model predicts a CO mixing ratio increasing with altitude in the troposphere and stratosphere, which is in agreement with~\citet{Bezard2002}, who imply both internal and external sources and show similar CO behaviour.
This illustrates the need to include external sources of CO to correctly explain its mixing ratio in the upper atmosphere, a phenomenon observed in Jupiter's atmosphere that is not the focus of this paper.
When comparing the two models, it is also worth mentioning that VULCAN does include benzene which ARGO does not.
Last, the model by~\citet{Tsai2021} extends to 5,000 bar whereas our model ends at 1,000 bar.
However, the atmosphere is in equilibrium even at 1,000 bar in our model and therefore the choice of lower boundary should not influence the result.

\begin{figure*}
    \centering
    \includegraphics[width=\textwidth]{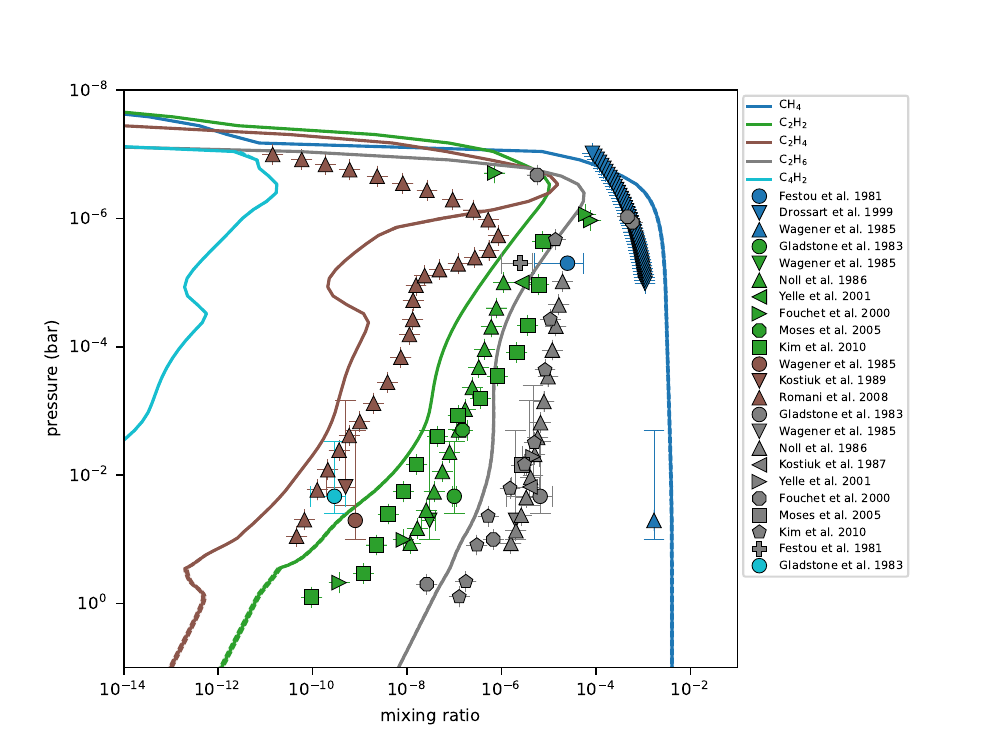}
    \caption{Mixing ratios of \ce{CH4}, \ce{C2H2}, \ce{C2H4}, \ce{C2H6} and \ce{C4H2} as a function of pressure compared to observations. Our results are shown in solid line with dashed line margins. Observational data were compiled from~\citet{Festou1981,Drossart1999,Wagener1985,Gladstone1983,Noll1986,Yelle2001,Fouchet2000,Moses2005,Kim2010,Kostiuk1987,Kostiuk1989,Romani2008}.}
    \label{fig:J135_as_Rimmer2016_fig_12_vs_obs}
\end{figure*}

The current model was also compared to observational data as shown in Figure~\ref{fig:J135_as_Rimmer2016_fig_12_vs_obs}.
Comparison with observational data shows that our model reproduces the general behaviour reasonably well.
Specifically, the predicted \ce{CH4} mixing ratio matches observations by~\citet{Drossart1999} and~\citet{Wagener1985}, although data from~\citet{Drossart1999} suggest a higher drop-off in the methane mixing ratio, matching the model of~\citet{Hu2021} better.
The value retrieved by~\citet{Festou1981} is about two orders of magnitude lower, but that paper uses a different pressure-temperature profile which disagrees with currently accepted profiles.
The mixing ratio profiles of \ce{C2H2}, \ce{C2H4} and \ce{C2H6} from our model are generally slightly lower than the observed mixing ratios.
Again, we attribute this fact mostly to the fact that the retrievals in the referenced papers use different and often simpler pressure-temperature profiles and/or simpler chemical models to infer the mixing ratios.
Interestingly, the observational data suggest a more linear increase in mixing ratio of the hydrocarbon species between 1 bar and the species' respective drop-off level, which agrees with our model very well (in contrast, e.g. to the former model in~\citet{Rimmer2016}.
Therefore, even though our absolute value mixing ratios are slightly different (a value which is sensitive to the lower boundary conditions, as discussed throughout this paper), the general behaviour is described very well.
The predicted mixing ratio of \ce{C4H2} differs from our model by about 4 orders of magnitude.
One explanation for this could be that the C4 species in our model act as terminating species and are therefore likely inaccurate.
We would like to point out at this point that the purpose of this model is to capture the general features of Jupiter's chemistry and possibly other Jovian planets and make predictions that are key for the explanation of some observable effects in the atmosphere, rather than formulation of a precise model of the current state of Jupiter.

\begin{figure}
    \centering
    \includegraphics{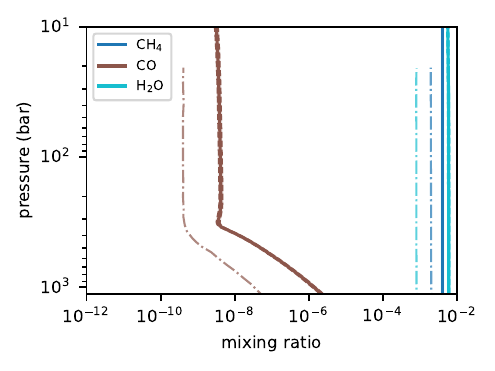}
    \caption{\ce{CO}, \ce{CH4} and \ce{H2O} mixing ratios in the Jupiter's deep atmosphere as a function of pressure (solid line with dashed line margins). The data are compared with results from~\citet{Visscher2010} (dash-dot line).}
    \label{fig:J135_as_Visscher2010_fig3}
\end{figure}

Next, our results were compared to  deeper atmospheric model by~\citet{Visscher2010}.
Figure \ref{fig:J135_as_Visscher2010_fig3} shows the vertical profiles of \ce{CO}, \ce{CH4} and \ce{H2O} plotted as in the compared paper.
Our model shows a similar general behaviour of the \ce{CO} mixing ratio.
However, we predict a quenched mixing ratio of CO $\sim$4.2 ppb, while the authors predict $\sim$0.4 ppb - an order of magnitude lower amount.
\citet{Visscher2010} use a water enrichment 2-4 times solar abundance and Kzz = $1 \times 10^{8}$~cm$^{-2}$\,s$^{-1}$ in their model to best reproduce the observed \ce{CO} mixing ratio at the 6 bar level of Jupiter~\citep{Bezard2002}.
Notably, this observed value was measured in a hot spot and the authors imply both internal and external sources of \ce{CO} with significant uncertainties in their magnitude, whereas our model is not tailored for the conditions in the hot spot.
One difference is that our model contains more water in the deep atmosphere (from the equilibrium calculations) than the model of~\citet{Visscher2010}, which may increase the mixing ratio of CO.
The difference in the deep amount of \ce{H2O} is also shown in the plot.
As visible from the lower boundary of the CO mixing ratio in our Figure \ref{fig:J135_as_Visscher2010_fig3}, which contains 5\% less water than the main calculation (lower margin), reducing the amount of deep water significantly reduces the predict amount of produced CO.
Indeed,~\citet{Visscher2010} use their model to constrain the amount of \ce{H2O} such that it produces the observed \ce{CO} mixing ratio, as \ce{H2O} is the main oxygen-containing specie in the Jovian deeper atmosphere.
Other reasons for the discrepancy may include a different treatment of the eddy diffusion, as even~\citet{Visscher2010} show in their paper that the value and hence the treatment of eddy diffusion may change the \ce{CO} abundance by as much as one order of magnitude (see their Figure 4).
We therefore predict a CO quench level at $\sim$318~bar and the quenched mixing ratio $4.2 \times 10^{-9}$.

\begin{figure*}
    \centering
    \includegraphics[width=\textwidth]{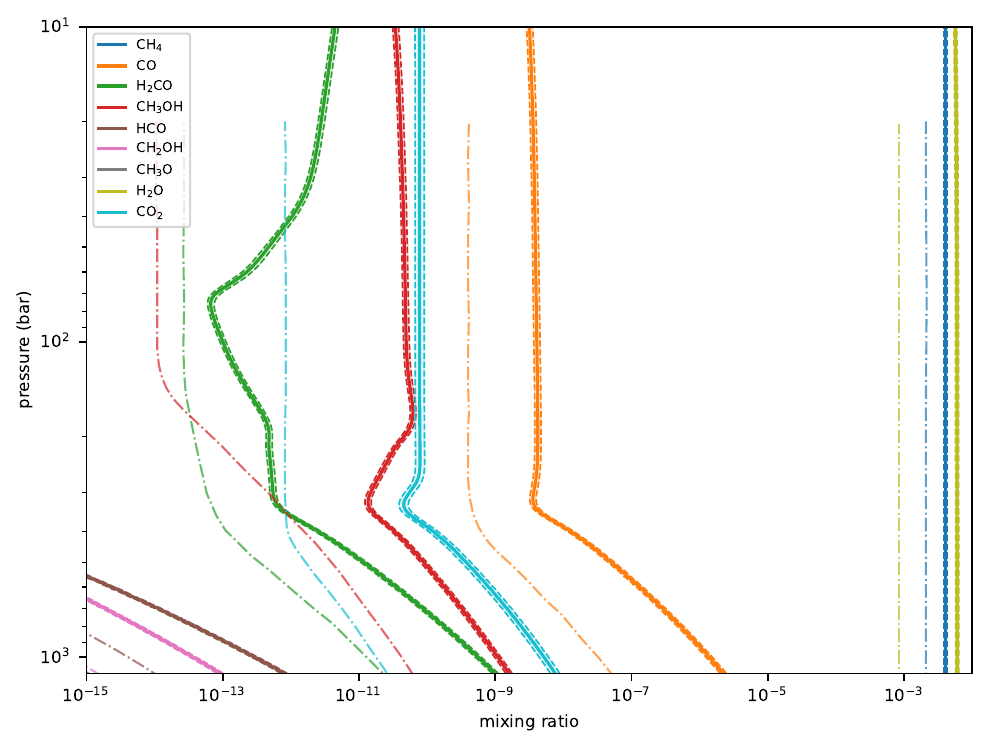}
    \caption{Mixing ratios of oxidized carbon-bearing species in the deep atmosphere (solid line with dash line margins) as a function of pressure. The data are compared with results of~\citet{Visscher2010} (dash-dot line).}
    \label{fig:J135_as_visscher2010_fig6}
\end{figure*}

Next, in Figure \ref{fig:J135_as_visscher2010_fig6}, we plotted the vertical profiles of oxidized carbon gases similar to what~\citet{Visscher2010} do in their Figure 6.
Once again, our models predicts a very similar behaviour of these species to what the model of~\citet{Visscher2010} predicts.
Notably, since our model contains more oxygen in the lower atmosphere, the corresponding amounts of not only water and \ce{CO}, but also all the other species (\ce{H2CO}, \ce{CH3OH}, \ce{CHO}, \ce{CH2OH}, \ce{CH2OH}, \ce{CH3O}, and \ce{CO2} are approximately one order of magnitude higher.
Our model, however, predicts quenching levels of \ce{CO}, \ce{CO2}, and \ce{H2CO} at basically exactly the same level as~\citet{Visscher2010}, which suggests that our description of the chemistry, pressure, temperature and eddy diffusion is accurate or similar to~\citet{Visscher2010} and that the changes might be attributed to the difference in the amount of oxygen.
Interestingly, our predicted quenched amount of \ce{CH3OH} is around $1.4 \times 10^{-11}$ and the quench level is $\sim$320~bar, as compared to $1.1 \times 10^{-14}$ at 110 bar by ~\citet{Visscher2010}.
Similarly to their paper, \ce{CH3OH} formation is dominated by reaction 
\begin{equation}
    \ce{CH3 + H2O -> CH3OH + H}
\end{equation}
below the quench level.
However, our model shows that the quench level occurs at around 320~bar, where the formation is dominated by reaction 
\begin{equation}
    \ce{CH3O + H2 -> CH3OH + H}
\end{equation}
and balanced by destruction from reaction 
\begin{equation}
    \ce{CH3OH + H -> CH3 + H2O}
\end{equation}
Above that, we observe an increasing mixing ratio of methanol, again from reaction 
\begin{equation}
    \ce{CH3 + H2O -> CH3OH + H}
\end{equation}
and then a stable quenched mixing ratio above 170~bar.
One possible difference in the models are different rate constants for the reactions in question.
Another possible explanation is a different vertical resolution of the model.
Our model uses a set vertical resolution 1~km, while~\citet{Visscher2010} use resolution of "at least 20 altitude levels per scale height".
At $\sim$450~bar, the scale height is $\sim$144~km, which means that with our resolution, we have 144 altitude levels per scale height.

\section{Ammonia and ammonium hydrosulfide} \label{sec:NH3_NH4SH}
The model was used to explore the chemistry of nitrogen in the Jupiter's deeper atmosphere, the troposphere and the stratosphere.
\begin{figure*}
    \centering
    \includegraphics[width=\textwidth]{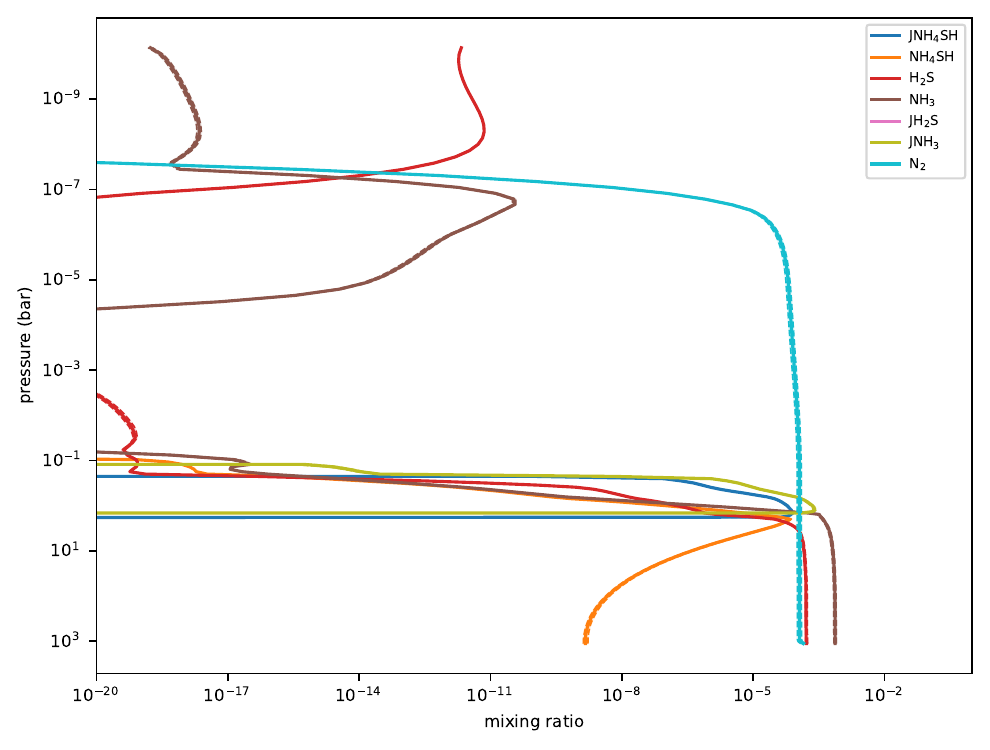}
    \caption{Mixing ratios of nitrogen bearing species on Jupiter as a function of pressure. Species denoted JX are condensed species.}
    \label{fig:J135_NH4SH}
\end{figure*}
The model contains the presence of ammonium hydrosulfide, a compound likely present on Jupiter and contributing to its orange-ish colour~\citep{Loeffler2018}.
As Figure \ref{fig:J135_NH4SH} shows, both \ce{NH4SH} and \ce{NH3} condense at a similar level around 1 bar and form a layer of cloud the reaches up to 0.1 bar.
Notably, \ce{H2S} does not condense by itself, but is consumed to produce \ce{NH4SH}.
Its mixing ratios therefore also drops sharply at this level, but the condensate does not appear.
Since \ce{H2S} is the major sulfur-bearing species in the deeper atmosphere and since absolute majority of it is consumed to produce \ce{NH4SH}, most of the sulfur is effectively trapped below the \ce{NH4SH} condensation level, although some \ce{H2S} is again photochemically formed in the upper stratospheric layers around $10^{-7} - 10^{-9}$ bar.
On the contrary, although ammonia condenses at the $10^0 - 10^1$~bar atmospheric level, it begins to photochemically form again at $5 \times 10^{-5}$~bar and reaches as much as $3.5 \times 10^{-11}$ mixing ratio at $2 \times 10^{-7}$~bar before being finally photochemically removed from the atmosphere.
The temperatures and mixing ratios of both \ce{H2S} and \ce{NH3} are sufficiently low in the stratosphere that no formation of \ce{NH4SH} is observed.
Similar reasoning is also employed in~\citet{Hu2021}.

\citet{Bolton2017} present \ce{NH3} mixing ratio map constructed from the initial results from the Juno spacecraft.
As they show in their Figure 3, ammonia abundance at $10^2$~bar is around 340-360~ppm across all latitudes and this abundance decreases with the decreasing pressure up until its condensation level.
The decrease is different for different latitudes.
Near the equator, the mixing ratio is rather constant and explained by rapid mixing with deeper layers, pointing to and effect similar to the Hadley cell.
At higher latitudes, the decrease in ammonia mixing ratio is slower and smoother.
In all cases, condensation occurs around 0.7~bar.
Our model predicts ammonia mixing ratio at $10^2$~bar 740~ppm, but rapidly decreasing with the decreasing pressure, down to 320~mbar at 1.55~bar where condensation starts to occur.
The behaviour of our model is therefore qualitatively similar to the observational data, although the condensation levels and the pre-condensation amounts are slightly different.

The most likely explanation is that the thermal profile needs to be updated down do deeper levels, although the measured Galileo probe data extend down to 6.7~bar.
Recently,~\citet{Gupta2022} have shown that Voyager radio occultation measurements give temperature at 1 bar up to 4~K different to the Galileo data with a possible variation as much as 7~K.
Differences in thermal profile would influence both the differences in ammonia abundance and condensation level, although chemical abundances at larger pressures would not be significantly affected~\citep{Rensen2023}.
Another possible explanation for the higher amount of ammonia is the fact that some reactions in the ARGO model are not reversed, as mentioned above.
A third possible explanation is that the deep atmosphere amount of N in this model is too high.

As discussed, the condensation level is predicted slightly deeper than was observed.
Condensation in this model is governed by the Antoine equation.
\begin{figure}
    \centering
    \includegraphics{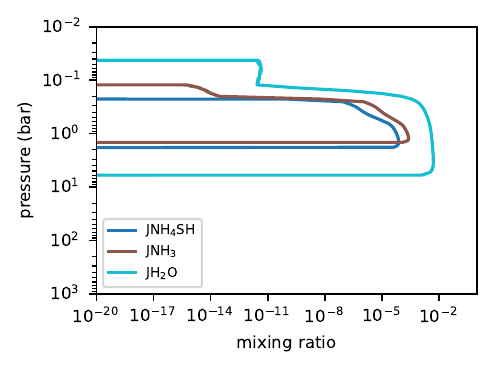}
    \caption{Condensation of the three major condensable species \ce{NH3}, \ce{H2O} and \ce{NH4SH} in our model. The plot shows mixing ratios as a function of pressure and JX signifies condensed species.}
    \label{fig:J135_condensation}
\end{figure}
As Figure \ref{fig:J135_condensation} shows, our model predicts condensation of ammonia and water at slightly higher pressure compared to~\citet{Rensen2023}.
This may be caused by the simplistic treatment of condensation in this model~\citep{Rimmer2016}.
The condensation chemistry on Jupiter is complicated to simulate.
The condensate clouds probably include not only pure ices of \ce{H2O}, \ce{NH3} and \ce{NH4SH}, but also ice mixtures of \ce{NH3} and \ce{H2O}, of which there are several types.
A layer of liquid \ce{H2O} is also expected~\citep{Atreya1999}.
In general, however, going up the atmosphere with an ever-decreasing pressure, water is the first to condense of these compounds, followed by \ce{NH4SH} and then \ce{NH3}~\citep{Rossow1978}.
Our model captures this order, although the condensation levels are very sensitive to the eddy diffusion coefficient. We have explored this sensitivity. In addition, icy mixtures are not accounted for, and the liquid phase and solid phase are not distinguished.
\citet{Bolton2017} explains the high abundance of ammonia near the equator by rapid mixing with the deeper atmosphere.
A last, most exciting option is that there is an unknown chemical process in the atmosphere, which could explain the abundance.

\begin{figure}
    \centering
    \includegraphics{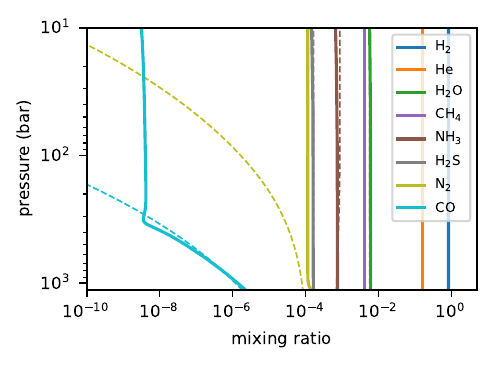}
    \caption{Mixing ratios of major component gases in Jupiter's deep atmosphere and troposphere. Our data are plotted in solid line and compared to~\citet{Rensen2023} (shown in dashed line).}
    \label{fig:J135_as_Rensen2023_fig2}
\end{figure}

\section{Deep tropospheric chemistry} \label{sec:deep troposphere}
Figure \ref{fig:J135_as_Rensen2023_fig2} shows the mixing ratios in Jupiter's deep atmosphere up to the troposphere.
In comparison with~\citet{Rensen2023}, our model predicts similar amounts of \ce{H2}, \ce{He}, \ce{H2O}, \ce{H2S}, \ce{CO} and \ce{CH4} - the main carriers of H, O, S and C in the deeper atmosphere.
The species behaviour in general is similar although the mixing ratios are slightly different.
Note that for graphical clarity, we did not plot the maximum and minimum values of our model.

There is a significant difference in the mixing ratio of \ce{N2}.
Our model predicts a relatively constant value of 11 ppm from the bottom of the atmosphere up to $10^{-6}$~bar where it starts to drop off.
On the other hand,~\citet{Rensen2023} show a gradual decrease of \ce{N2} mixing ratio from $\sim 8.5 \times 10^{-4}$ at 1000~bar to below $10^{-10}$ at a few bar level.
This is due to the fact that the authors use a thermochemical equilibrium model.
However, quenching maintains a relatively constant amount of nitrogen even in the higher atmosphere.
Quenching is a process where the timescale for vertical transport becomes important, because as temperature decreases, thermochemical reactions slow down to the point where vertical transport is faster, species stop reacting and remain at a constant level throughout the atmosphere above.
This is exactly the case of \ce{N2}, which gradually stops converting to \ce{NH3} (as it would in equilibrium) and remains at a fixed value $\sim$11ppm up until $10^{-6}$~bar where it begins to photochemically decompose, making it an important disequilibrium species in the Jupiter's atmosphere.
Similar behaviour for \ce{N2} was described by~\citet{Moses2010}.
\begin{figure*}
    \centering
    \includegraphics[width=\textwidth]{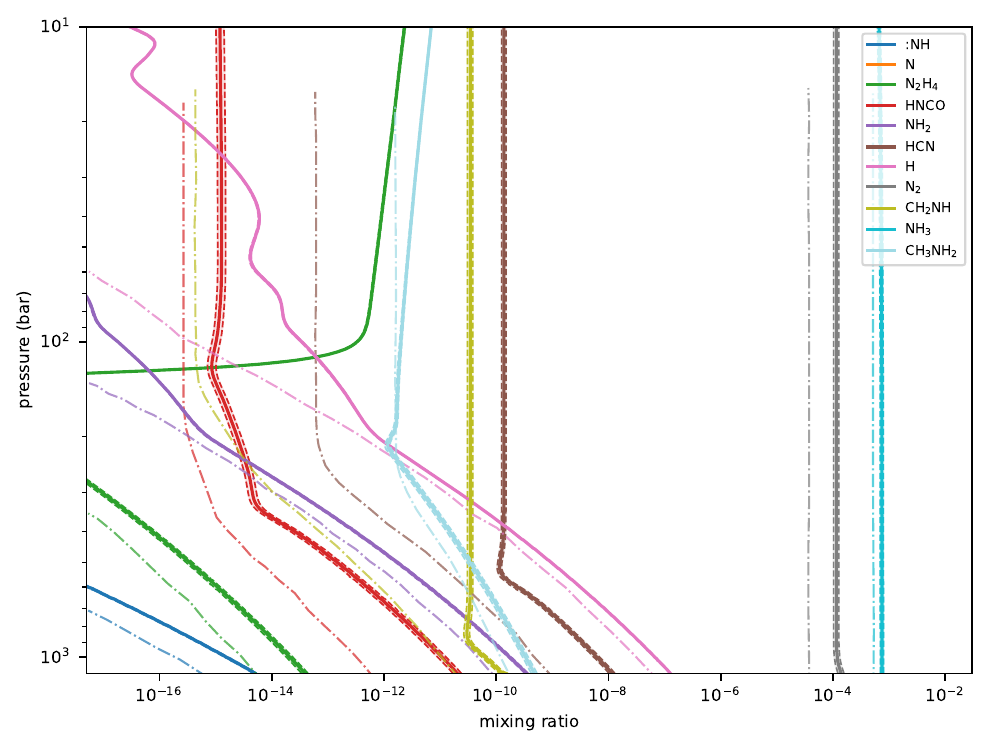}
    \caption{Mixing ratios of nitrogen-containing species as a function of pressure in Jupiter's deep atmosphere. Our model results are plotted in solid line with dashed line margins. The results are compared to~\citet{Moses2010}, Figure 4, shown here in dash-dot line.}
\end{figure*}
The figure shows mixing ratios of selected nitrogen species.
Comparison of the original figure in~\citet{Moses2010} and our result shows similar behaviour for the observed species (and similar results are obtained by \citet{Wang2016} as well).
One notable difference is the offset of quench levels, which appear at higher pressures in our model, resulting in increased quenched abundances of \ce{HCN}, \ce{HNCO} and \ce{CH2NH}.
The most likely explanation for this difference is a difference in thermal profile of the atmosphere used for the respective models~\citep{Hu2021}.
Our model also includes more \ce{N2} and \ce{NH3}.
The overall greater abundance of available nitrogen also contributes to the higher amount of all nitrogen species.
The largest observed difference is in the amount of \ce{HCN}, of which~\citet{Moses2010} predict a quenched abundance $6 \times 10^{-14}$ and our model predicts $1.4 \times 10^{-10}$.
The chemistry of HCN is in detail discussed in the following section.

\section{Hydrogen cyanide} \label{sec:HCN}
One interesting outcome of our model is the formation of \ce{HCN} in the stratosphere.
\ce{HCN} on Jupiter is a molecule of great interest, because it is a molecule which plays a role in prebiotic chemistry~\citep{Sutherland2016,Ferus2017}, is a relevant tracer for nitrogen chemistry, can potentially be used to trace impacts~\citep{Rimmer2020}, and because its polymers may be contributing to the planet's colour~\citep{Woeller1969,Matthews1991}.

\begin{figure}
    \centering
    \includegraphics{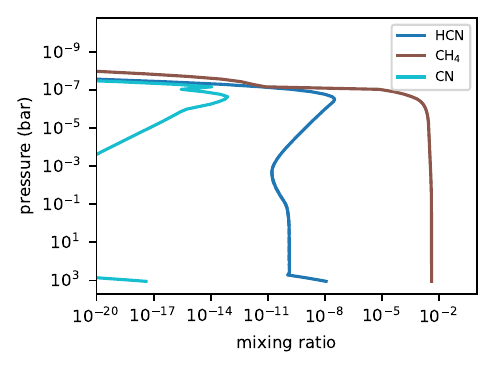}
    \caption{Atmospheric mixing ratios of \ce{HCN}, \ce{CN} and \ce{CH4} from our model as a function of pressure. And increase in \ce{HCN} mixing ratio is clearly visible.}
    \label{fig:J135_HCN}
\end{figure}
\ce{HCN} has been detected on Jupiter.
\citet{Tokunaga1981} detected the R-branch lines of the $\nu_{2}$ fundamental band of \ce{HCN} from which they inferred a \ce{HCN} column density $5 \times 10^{-3}$~cm-am, or mixing ratio $2 \times 10^{-9}$ from 0.6~bar to the top of the atmosphere.
Later,~\citet{Weisstein1996} observed \ce{HCN} on Jupiter and put an upper limit 0.3~ppb if distributed with a constant mixing ratio.
The authors also claim that if condensation was considered in their work, the upper limit would increase to 2~ppb.
Most recently, these upper limits were refined by~\citet{Davis1997} to 0.93~ppb if \ce{HCN} condenses in the upper troposphere or 0.16~ppb if it is uniformly mixed throughout the troposphere and stratosphere.
In 1994, the impact of comet Shoemaker-Levy 9 induced large changes in the Jupiter's atmosphere concerning \ce{HCN}.
\ce{HCN} quickly spread through the atmosphere which remains altered by the impact to this day~\citep{Moreno2003,Lellouch2006}.

\citet{Moses2010} model Jupiter's tropospheric nitrogen chemistry in an attempt to explain its low observational upper limit.
At the same time, more recently,~\citet{Hu2021} predict low ($~10^{-9}$) mixing ratios of \ce{HCN} throughout Jupiter's atmosphere.
They observe formation of \ce{from} \ce{NH3} and \ce{CH4} and give small mixing ratios mostly due to the fact that photolysis of \ce{NH3} and \ce{CH4} occurs at well separated pressure levels and transport is limited by the cold trap near the tropopause.

Investigation of \ce{HCN} chemistry in our model revealed six distinct regions with its specific major \ce{HCN} production and destruction rates.
The mechanisms were determined by looking for the fastest reaction at each level which forms HCN and subsequent tracing of the fastest formation and destruction reactions of its precursors throughout the STAND network.
\begin{figure}
    \centering
    \includegraphics{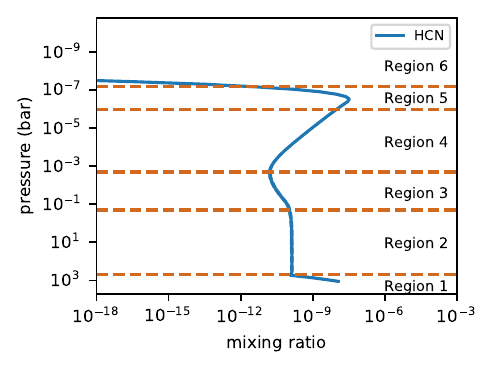}
    \caption{HCN mixing ratio in our model as a function of pressure with six marked regions of different dominant chemistry.}
    \label{fig:HCN_six_regions}
\end{figure}
Figure \ref{fig:HCN_six_regions} shows these six different regions.
The first region, deepest in the atmosphere is governed by chemical equilibrium.
In this region, \ce{HCN} is formed from \ce{CO} and \ce{NH3} by the following mechanism.
\begin{align}
    &\ce{NH3 -> NH2 + H} \nonumber\\
    &\ce{CO + NH2 -> HNCO + H} \nonumber\\
    &\ce{HNCO + H -> OH + HNC} \nonumber\\
    &\ce{HNC -> HCN} \nonumber\\
    &\ce{H + OH -> H2O} \nonumber \\
    &\rule[3mm]{52mm}{.8pt} \nonumber \\
    \mathrm{Net:} \: &\ce{CO + NH3 -> HCN + H2O}
\end{align}
\ce{HCN} in this region is mainly formed from its stable precursors, but the formation is slower as one moves up the atmosphere, resulting in a decrease in the mixing ratio.

The second region of \ce{HCN} chemistry lies approximately between 517~bar and $2 \times 10^{-1}$~bar.
In this region, the \ce{HCN} mixing ratio remains mostly constant due to the fact that reactions are quenched and  the \ce{HCN} is transported upwards without further reactions.
This also implies that the lower boundary of this region, i.e. around 517~bar, is the \ce{HCN} quenching level.
The main production of HCN is still from HNC, but the net production rate is $\sim 1 \times 10^{-3}$ molec\,cm$^{-2}$\,s$^{-1}$, which is much lower than the vertical transport rate.
Even more, the production of HNC from any source is several orders of magnitude lower than that.

The third distinct region lies between $2 \times 10^{-1}$~bar and $2 \times 10^{-3}$~bar.
Here the \ce{HCN} mixing ratio decreases with altitude faster than in Region 2.
\ce{HCN} is removed dominantly by the following mechanism.
\begin{align}
    &\ce{HCN ->[\textit{h\nu}] H + CN} \nonumber\\
    &\ce{CN + CH4 -> CH3CN + H} \nonumber\\
    &\ce{CH3CN + CH3 -> CH2CN + CH4} \nonumber\\
    &\ce{2H -> H2} \nonumber \\
    &\rule[3mm]{55mm}{.8pt} \nonumber \\
    \mathrm{Net:} \: &\ce{HCN + CH3 -> CH2CN + H2}
\end{align}
The first reaction is a photochemical splitting of HCN, so in this region, photochemistry already takes place.
\ce{CH3} in this region is produced mostly from \ce{CH2X} by the reaction
\begin{equation}
    \ce{CH2X + H2 -> CH3 + H}
\end{equation}
where \ce{CH2X} is a catch-all species for carbohydrates in the code, which is rigorous for carbon-containing compounds up to C2-C3 compounds.
In summary, then \ce{HCN} in this region is destroyed by photochemistry and carbohydrates.

The fourth region spans from $2 \times 10^{-3}$~bar to $1 \times 10^{-6}$~bar.
Although the HCN mixing ratio increases in this part of the atmosphere, the region sees photochemical HCN destruction as the dominant process:
\begin{align}
    &\ce{HCN ->[\textit{h\nu}] H + CN} \nonumber\\
    &\ce{CN + CH4 -> CH3CN + H} \nonumber\\
    &\ce{2H -> H2} \nonumber \\
    &\rule[3mm]{55mm}{.8pt} \nonumber \\
    \mathrm{Net:} \: &\ce{HCN + CH4 -> CH3CN + H2}
\end{align}
The net reaction is reaction with CH4 to produce \ce{CH3CN}, which accumulates in the atmosphere, but the first necessary step is photochemical destruction of HCN.

Next, between $1 \times 10^{-6}$ bar and $6.76 \times 10^{-8}$ bar lies a newly discovered region of HCN formation.
This is the fifth distinct region of HCN chemistry.
The formation proceeds through several reactions and again includes photochemistry and radical reactions:
\begin{align}
    &\ce{CH4 ->[\textit{h\nu}] CH2 + 2H}\nonumber\\
    &\ce{CH2 + H -> H2 + CH}\nonumber\\
    &\ce{CH + CH4 -> C2H4 + H}\nonumber\\
    &\ce{C2H4 + H -> C2H5}\nonumber\\
    &\ce{C2H5 + H -> 2CH3}\nonumber\\
    &\ce{N2 ->[\textit{h\nu}] 2N}\nonumber\\
    &\ce{2N + 2CH3 -> 2H2CN + 2H}\nonumber\\
    &\ce{2H2CN -> 2HCN + 2H}\nonumber\\
    &\ce{4H -> 2H2} \nonumber \\
    &\rule[3mm]{55mm}{.8pt} \nonumber \\
    \mathrm{Net:} \: &\ce{2CH4 + N2 -> 2HCN + 3H2}
\end{align}
The spare N and H radicals produced throughout the mechanism enter further reactions in the atmosphere and recombine.
The region also sees the formation of \ce{CH3CN} and \ce{HC3N}, whose mixing ratios don't exceed $3 \times 10^{-11}$ and $5 \times 10^{-10}$, respectively.
Their potential for haze formation is therefore quite limited, although this model does not include any haze formation routines.

Last, the production and destruction of HCN are very slow and balanced from the $6.76 \times 10^{-8}$ bar up to the top of the atmosphere.
As the atmosphere thins, \ce{H2} is photochemically split to H and upwards transport is quenched, the mixing ratio finally decreases.

Overall the upper atmospheric HCN mixing ratio reaches a peak 33 ppb at the $2.94 \times 10^{-7}$~bar level.
Even though the production region is quite narrow in the atmosphere, vertical transport causes the mixing ratio of HCN to increase from the $2 \times 10^{-3}$ bar.

\begin{figure}
    \centering
    \includegraphics{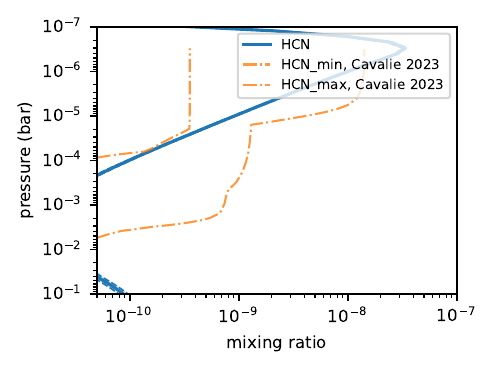}
    \caption{HCN mixing ratio in our model as a function of pressure from our model (blue) and as observed by \citet{Cavalie2023} (orange).}
    \label{fig:HCN_vs_obs}
\end{figure}

Recent observations of \ce{HCN} and \ce{CO} in the stratosphere of Jupiter by~\citet{Cavalie2023} retrieve a profile similar to our model.
A comparison is shown in Figure \ref{fig:HCN_vs_obs}.
Their profile is based on observations and predicts a uniform \ce{HCN} mixing ratio between 0.2 and 20~ppbv in the upper stratosphere and a sharp drop-off at $2^{+2}_{-1}$ and $0.04^{0.07}_{0.03}$~mbar.
The authors propose that heterogeneous chemistry bonds \ce{HCN} on aerosol particles, causing the observed depletion.
Our model does not include aerosol chemistry and would therefore not catch this effect.
Interestingly though, the predicted volume mixing ratios are of similar order of magnitude.
The pressure range in question covers part of Region 3 and part of Region 4 from our model.
In Region 3, our model also predicts depletion of \ce{HCN} by reaction with \ce{CH3} (as discussed above).
The change in the chemical regime occurs around 2~mbar, which roughly corresponds to observations by~\citet{Cavalie2023}, although the proposed mechanisms are different.
In Region 4, or above the $\sim$ 2~mbar pressure level,~\citet{Cavalie2023} predicts a uniform mixing ratio of \ce{HCN} while our model predicts \ce{HCN} formation by several mechanisms.
The resulting amounts are, however, of the same order of magnitude.

The earliest next mission to observe Jupiter and possibly test our prediction is the JUICE (Jupiter Icy Moons Explorer) mission by ESA.
This mission is primarily designed to study Jupiter’s icy moons, but its spectrometer MAJIS will observe Jupiter’s atmosphere as well.
The spectrometer will focus on improving the current Jupiter data by increasing the spatial resolution of the current observational datasets and will characterize both the atmospheric composition and dynamics~\citep{Poulet2024}.

\section{Conclusion} \label{sec: conclusion}
Jupiter's deeper atmosphere, troposphere and stratosphere are for the first time combined in a single 1D chemical kinetics model.
The model was benchmarked and stands relatively well in comparison to existing region-specific models in the literature.
The model code is based on the ARGO code published in~\citet{Rimmer2016}.
This current iteration of the code contains a new edition of the STAND reaction network with \ce{NH4SH} chemistry as well as updates to Antoine equation parameters for condensation of \ce{NH4SH} and \ce{H2S}.
We discuss throughout the paper that thermal profile and initial conditions at the lower boundary are the most important effects on the resulting mixing ratios and quench levels of atmospheric species.
The formation and condensation of \ce{NH4SH} from \ce{H2S} and \ce{NH3} traps the majority of sulfur at the lower levels of the atmosphere.
Both ammonia and \ce{NH4SH} condense and form cloud layer between 0.1 and 1 bar atmospheric level, likely contributing to the planet's orange-ish colour.
Comparison with recent observations by Juno shows that our model best represents behaviour of the mixing ratio of ammonia near equatorial latitudes, where rapid mixing and massive updrafts are the dominant atmospheric effect, as compared to other latitudes, where heterogeneous chemistry, global circulation and aerosols play a more important part.
By comparison with thermochemical equilibrium model by~\citet{Rensen2023}, we also show the importance of quenching for the atmosphere.
Our model predicts \ce{HCN} formation between $1 \times 10^{-6}$ and $6.76 \times 10^{-8}$~bar with a maximum mixing ratio of 33~ppb at $2.94 \times 10^{-7}$ and details the mechanism of its formation through radical chemistry and photochemistry.
This is an important prediction which can be tested by observations.

\section{Declaration of competing interest}
The authors declare that they have no known competing financial interests or personal relationships that could have appeared to influence the work reported in this paper.

\section{Data availability}
Data for this paper are available from the corresponding author upon reasonable request.

\section{Acknowledgements}
We thank the Department of Physics, Cavendish Laboratory, University of Cambridge for their start-up funding that provided support for this project. MF thanks the Czech Science Foundation project reg. no. 24-12656K. The authors acknowledge the assistance provided by the Advanced Multiscale Materials for Key Enabling Technologies project, supported by the Ministry of Education, Youth, and Sports of the Czech Republic. Project No. CZ.02.01.01/00/22 008/0004558, Co-funded by the European Union. We also greatly acknowledge the support of the regional collaboration with Valašské Meziříčí observatory provided by a grant of reg. no. R200402401. We gratefully acknowledge support of the development of planetary models with ESA PRODEX under contract PEA4000147310.
\appendix
\section{My Appendix}
\printcredits

\bibliographystyle{cas-model2-names}

\bibliography{references}

\end{document}